\documentclass[preprint,showpacs,preprintnumbers,amsmath,amssymb,aps,pre]{revtex4-1}

\usepackage{color}

\usepackage{graphicx}
\usepackage{dcolumn}
\usepackage{bm}

\begin{document}


\title{On the relevance of q-distribution functions:  The return time distribution of restricted random walker}

\author{Jaleh Zand$^{1,}$}
 \email{jaleh.zand13@imperial.ac.uk}
\author{Ugur Tirnakli$^{2,}$}
 \email{ugur.tirnakli@ege.edu.tr}
\author{Henrik Jeldtoft  Jensen$^{1,}$}
 \email{h.jensen@imperial.ac.uk}

\affiliation{
$^1$Centre for Complexity Science and Department of Mathematics, \\ 
Imperial College London, South Kensington Campus, London SW7 2AZ, UK\\
$^2$Department of Physics, Faculty of Science, Ege University, 35100 Izmir, Turkey\\
}

\date{\today}

\date{\today}

\begin{abstract}
There exist a large literature on the application of $q$-statistics to the out-of-equilibrium 
non-ergodic systems in which some degree of strong correlations exists.   
Here we study the  distribution of first return times to zero, $P_R(0,t)$, of a random walk on the set of integers $\{0,1,2,...,L\}$ with a position dependent transition probability given by $|n/L|^a$. We find that for all values of $a\in[0,2]$ $P_R(0,t)$ can be fitted by $q$-exponentials, but only for $a=1$ is $P_R(0,t)$ given exactly by a $q$-exponential in the limit $L\rightarrow\infty$. This is a remarkable result since the exact analytical solution of the corresponding continuum model represents $P_R(0,t)$ as a sum of Bessel functions with a smooth dependence on $a$ from which we are unable to identify $a=1$ as of special significance. However, from the high precision numerical iteration of the discrete Master Equation, we do verify that only for $a=1$  is $P_R(0,t)$  {\it exactly} a $q$-exponential and that a tiny departure from this parameter value makes the distribution deviate from $q$-exponential. Further research is certainly required to identify the reason for this result and also the applicability of $q$-statistics and its domain.
\end{abstract}

\maketitle

\section{Introduction}

It is well know that Boltzmann-Gibbs statistics is related to the exponential and Gaussian functions because these functions are the distributions corresponding to maximising the Boltzmann-Gibbs entropy.  On the other hand, in cases where ergodicity is broken say due to strong correlations or long range interaction exist Boltzmann-Gibbs statistics can break down. It has been suggested that for a very wide class of such situations including experimental \cite{expr1,expr2,expr3}, observational \cite{obs1,obs2,obs3} and examples of model systems \cite{model1,model2,model3,model4,model5}, the so-called $q$-statistics is the appropriate generalization of Boltzmann-Gibbs statistics \cite{tsallis88,tsallisbook}. In this case ordinary exponential and Gaussian functions are replaced by $q$-exponentials and $q$-Gaussians, defined respectively as,

\begin{equation}
\exp_q u =\left\{\begin{array}{ll}
                              \left[1+(1-q)u\right]^{1/(1-q)} \,\,\, ,&\,\,\, 
                              \mbox{$1+(1-q)u\geq 0$}\\
                              \,\,\,\,\,\,\,\,\,\,\,\,\,\,\,\,\,\,  0 \,\,\,\,\,\,\,\,\,\,\,\,\,\,\,\,\,\,\,\,\,\,\,\,\,\,\,\,\,\,\,\,\,,
                              &\,\,\,\,\,\,\,\,\,\,\,\,\,\,\,\,\,\, \mbox{else} \,
                                \end{array}
                           \right.
\label{qexp}
\end{equation}

and 

\begin{equation}
P(u)\propto \exp_q(-B u^2)
\label{qgaus}
\end{equation}
which maximise Tsallis entropy 
$S_{q} \equiv k \left(1- \sum_i p_i^q\right)/ \left(q-1\right)$ under appropriate conditions \cite{tsallis88,tsallisbook}.  Much of the evidence for this suggestion comes from often very good quality fitting of the $q$-functions to simulated or observed data combined with the appeal of the idea that the $q$-functions are the maximum distributions an appropriate entropy of presumably very broad applicability. This view point suggests that when dealing with systems that takes us beyond 
the realm of Boltzmann-Gibbs, the $q$-functions should be the generic distribution functions.  
Accordingly we would expect $q$-functions describe the system whenever the choice of systems 
parameters leads to non-Boltzmann-Gibbs behavior. Given this, one would furthermore expect that even 
if we are not {\em a priori} able to determine the specific value of the $q$ parameter 
in Eqns. (\ref{qexp}) and (\ref{qgaus})  we should be able to obtain the relevant value of $q$ by 
fitting since if $q$-statistics is the generic generalisation of Boltzmann-Gibbs statistics the family 
of $q$-functions should exhaust the functional forms distributions will assume. In order to understand 
better to which extend this attractive picture apply in reality, two of the authors together with 
C. Tsallis investigated a little while ago a one-dimensional Restricted Random Walk (RRW) 
model \cite{EPL}. The model consists of a random walker with a reflecting boundary condition on the set 
of integers $\{0,1,2,...,L\}$, where the probability to make a move is given by $g(n)=(n/L)^a$. 
The Master Equation for the model is straight forward to derive and to solve numerically by iteration 
to any desired numerical precision. We studied the  sum of the positions of the walk after $T$ steps.  
From very high precision numerics we concluded that in the limit $L\rightarrow \infty$ and for  
a certain scaling of $T$ and $L$ the distribution of this sum does not converge to the usual Gaussian 
given by the ordinary central limit theorem, but rather the limit distribution is a $q$-Gaussian. 
Surprisingly the $q$-Gaussian is only obtained for $a=1$. For $a\neq1$ neither a Gaussian nor a 
$q$-Gaussian appears to describe the limit distribution of the sum of positions. It is not 
immediately clear why the value $a=1$ for the exponent of the transition probabilities is a special case.

In order investigate how $a=1$ can be a singular value, we here investigate the distribution of first return times of the RRW. We solve the Master Equation for the probability to find the walker at position 
$n$ at time step $t$ analytically exactly in the continuum approximation and also solve numerically exactly 
the discrete version of the walker. From the analytic expression, which gives the distribution in terms of a Fourier-Bessel series, we are unable to identify anything singular about the value $a=1$. However, 
the numerical solution indicates that indeed for $a=1$ the return distribution is given by a 
$q$-exponential in the limit $L\rightarrow\infty$, whereas for $a\neq1$ the distribution is close 
but not exactly equal to any $q$-exponential.

 In the remainder of the paper we present firstly the analytic solution to the Master Equation for 
 the RRW model and investigate the first return distribution. We use the analytic solution presented 
 here to investigate the nature of the dependence of the distributions on $a$, and we then assess 
 the relevance of the $q$-statistics by numerically approaching the same problem. We close with a 
 discussion of our results and their implications for how $q$-functions may relate to non-Boltzmann-Gibbs 
 statistics.

\section{The Restricted Random Walker}
We consider a one-dimensional RRW model. We let $n$ be the position of this model on the set 
$\{0,1,2,...,L\}$, and $s$ be the discrete time variable. The walker is confined to the integers 
between 0 and $L$, and therefore $n \in \{0,1,2...,L\}$. The motion of the walker is controlled by 
the following process:
\begin{equation}
 n_{s+1}=
\left \{
  \begin{tabular}{ccc}
  $n_s+1$ with probability $g(n)/2$\\
   $n_s-1$ with probability $g(n)/2$\\
   $n_s$ with probability $1 - g(n)$ 

  \end{tabular}
\right \}
\label{process}
\end{equation}
where

\begin{equation}
g(n)=\left|\dfrac{n}{L}\right|^a~,~~ a \in (0,2)
\label{Eqg}
\end{equation}
Noting that the process for a normal random walker requires that $g(n)=1$ so that the walker moves to 
the left or right with a probability of $1/2$ at each time-step.

\section{The discrete Master Equation for the distribution}
The Master Equation for the distribution is given by
\begin{equation}
\begin{split}
P_{n}(n,s+1)=&P_{n}(n,s)+\dfrac{1}{2}g(n-1)\\
&\times P_{n}(n-1,s)+\dfrac{1}{2}g(n+1)\\
&\times P_{n}(n+1,s) -g(n)P_{n}(n,s)  .
\label{EqPD}
\end{split}
\end{equation}
We note that the Master Equation given in Eq.~(\ref{EqPD}) is valid for the bulk sites.  At $n=L$ we 
have a reflective boundary condition, while at $n=0$ we have an absorbing boundary condition. 
The probability of the return for the RRW at $n=0$, is therefore as follows:

\begin{equation}
P_{R_D}(0,s+1)=P_{n}(0,s+1)-P_{n}(0,s)
\label{EqR}
\end{equation}

\section{The continuous equation for the distribution}


\subsection{The Master Equation in continuous form}\label{continuous}

In order to obtain the continuous form of the Master Equation we introduce  $\Delta x=\dfrac{1}{L}$, 
i.e., $x=n\Delta x\in\left\{0,\dfrac{1}{L},\dfrac{2}{L},...,1\right\}$.  
Similarly we scale time and choose a time increment $\Delta t =\dfrac{1}{L}$ or $t=\dfrac{s}{L}$. 
This ensures that  the continuum approximation becomes exact in the limit $L\rightarrow\infty$. 
Finally we replace $g(n)$ by $g(x)$, where $g(x)=\left|x\right|^a$.  

As $L\rightarrow\infty$, the Master Equation takes the form of the following continuum diffusion equation
\begin{equation}
\dfrac{\partial P(x,t)}{\partial t} =\dfrac{1}{2L}\,\dfrac{\partial^2}{\partial x^2}\left[x^a\,P(x,t)\right]
\label{EqPC}
\end{equation}
A similar problem was studied in \cite{ben-Avraham1990}. Using the separation of variables the solution 
for $P(x,t)$ can be found as

\begin{equation}
P(x,t)=\sum\limits_{\lambda} A_{\lambda}\,P_{\lambda}(x)\,\exp\left(-\dfrac{\lambda^2}{8L}t\right)  .
\label{EqSolG}
\end{equation}
Assuming $Q_{\lambda}(x)=x^a\,P_{\lambda}(x)$, the differential equation in space becomes

\begin{equation}
\dfrac{d^2}{d x^2}Q_{\lambda}(x) +\dfrac{\lambda^2}{4x^a}\,Q_{\lambda}(x)=0 .
\label{EqMain}
\end{equation}
The general solution of this equation can be written in terms of Bessel and Neumann 
functions \citep[p.~362, 9.1.51]{Reference1}:

\begin{equation}
\begin{split}
Q_{\lambda}(x)=&A\,\sqrt x\,J_{\mbox{\scriptsize 1/(2-{\it a})     }}  
\left(\dfrac{\lambda\, x^{1-a/2}}{2-{\it a}}\right)\\
&+B\,\sqrt x\,Y_{\mbox{\scriptsize 1/(2-{\it a})     }}  \left(\dfrac{\lambda\, x^{1-a/2}}{2-a}\right) .
\label{EqAbrom}
\end{split}
\end{equation}

The boundary conditions are the following: 
At $x=0$ given that $P_{\lambda}(x)$ is finite, $Q_{\lambda}(x=0)$ is zero.  At $x=1$ we have a 
reflective boundary, therefore the current $J$, at $x=1$ is zero i.e. $J(x,t)\mid_{x=1}=0$.  
Further we know that $\dfrac{\partial P}{\partial t}=-\dfrac{\partial J}{\partial x}$.  
Using this fact together with Eq.~(\ref{EqPC}) we can conclude that 
$J(x,t)=-\dfrac{1}{2L}\,\dfrac{\partial }{\partial x}\left[x^a\,P(x,t)\right]$.  
We know that if $\dfrac{d}{dx}Q_{\lambda}(x)\mid_{x=1}=0$, this condition is satisfied.

The boundary condition at  $x=0$ implies that $B=0$.  The boundary condition at $x=1$ implies that 
$\dfrac{\lambda}{2-a}$ has to be the zero root of the Bessel function of order 
$\left\{\dfrac{1}{2-a}-1\right\}$.  
Therefore $\dfrac{\lambda_n}{2-a}=j_{\mbox{\scriptsize 1/(2-{\it a}) -1    },n}$, where 
$j_{\mbox{\scriptsize 1/(2-{\it a}) -1    },n}$ is the $n$th zero root of the Bessel function of 
order $\left\{\dfrac{1}{2-a}-1\right\}$.

So we arrive at the follwoing solution for $P(x,t)$  
\begin{equation}
\begin{split}
P(x,t)=&\sum\limits_{n=1}^\infty   A_{n}\,\dfrac{J_{\mbox{\scriptsize 1/(2-{\it a})}} 
\big(j_{\mbox{\scriptsize 1/(2-{\it a}) -1    },n}\,x^{1-a/2}\big)  }{x^{a-1/2}}\\
&\times\exp\Bigg(-\dfrac{{(2-a)}^2\,j_{\mbox{\scriptsize 1/(2-{\it a}) -1    },n}^2}{8L}\,\,t\Bigg)  .
\label{EqGeneral}
\end{split}
\end{equation}
%
For the special case of $a=1$, it turns out that

\begin{equation}
P(x,t)=\sum\limits_{n=1}^\infty   A_{n}\,\dfrac{J_1 \big(j_{0,n}\,\sqrt x\big)  }{\sqrt x}\,
\exp\Big(-\dfrac{j_{0,n}^2}{8L}\,\,t\Big)  .
\label{EqGena1}
\end{equation}

\subsection{The solution of the coefficient $A_n$}
Our initial condition for the problem is that at $t=0$, all walkers are placed at $n=1$.  
We attempt to solve coefficient $A_n$ for this initial condition.  
At $t=0$ the distribution is as follows:

\begin{equation}
P(x,0)=\sum\limits_{n=1}^\infty   A_{n}\,\dfrac{J_{\mbox{\scriptsize 1/(2-{\it a})}} 
\big(j_{\mbox{\scriptsize 1/(2-{\it a}) -1    },n}\,x^{1-a/2}\big)  }{x^{a-1/2}}   .
\label{EqGeneralt0}
\end{equation}
To solve coefficients $A_n$ we refer to the Fourier-Bessel series \cite[Chap. 8]{Bessel}.  
We note that we have a slightly different problem here.  We therefore present a modified version 
of the  Fourier-Bessel series below.

We note that Eq.~(\ref{EqMain}) is a standard Sturn-Liouville equation \cite{Sturn}, and as a 
result its eigenfunctions are orthogonal.  
Therefore if $a$ and $b$ are different zeros of the Bessel function of order $\nu-1$, then 
$\int J_\nu(ax)\, J_\nu(bx)\,x\,dx=0$. If $a\rightarrow b$ the solution to this integral is 
$J_{\nu}^2(a) /2$ \cite{Maple}. Hence one can easily write

\begin{equation}
\int_0^1 J_\nu(j_{\nu-1,m}\,x)\, J_\nu(j_{\nu-1,n}\,x)\,x\,dx=\dfrac{\delta_{m,n}}{2}
\left[J_{\nu}^2(j_{\nu-1,m})\right]
\label{IntegG2}
\end{equation}
where $\delta_{m,n}$ is the Dirac delta function.\\

Using the above result, we obtain
\begin{equation}
A_n=\dfrac{2(1-a/2)}{J_{\mbox{\scriptsize 1/(2-{\it a})}}^2 \big(j_{\mbox{\scriptsize 1/(2-{\it a}) -1    },n}\big)\,
\sqrt L}J_{\mbox{\scriptsize 1/(2-{\it a}) }}\Bigg(\dfrac{j_{\mbox{\scriptsize 1/(2-{\it a}) }-1,n}}{L^{1-a/2}}\Bigg)  .
\label{An1}
\end{equation}
For the special case of $a=1$, this result reads 

\begin{equation}
A_n=\dfrac{1}{J_1^2 (j_{0,n})\,\sqrt L}J_1\Bigg(\dfrac{j_{0,n}}{\sqrt L}\Bigg)  .
\label{An2}
\end{equation}

We now consider the asymptotic behaviour for large $t$. Using the asymptotic expansion for 
$J_{\nu}(z)$ \citep[p.~199, 7.21]{Bessel}, as $n\rightarrow \infty$, $A_n$ 
is asymptotically equal to

\begin{equation}
\begin{split}
A_n\sim& \sqrt {2\pi\,j_{\mbox{\scriptsize 1/(2-{\it a}) }-1,n}}\,\Big(1-\dfrac{a}{2}\Big)\,L^{-a/4}\\
&\times\cos \left(\dfrac{j_{\mbox{\scriptsize 1/(2-{\it a}) }-1,n}}{L^{1-a/2}}-\dfrac{\pi}{2(2-a)}-\dfrac{\pi}{4}\right) .
\label{asympAn}
\end{split}
\end{equation}
This implies that as long as $\dfrac{t}{L}$ is not too small, the higher terms in the sum in 
Eq.~(\ref{EqGeneral}) can be ignored and $P(x,t)$ will decay exponentially for $t>L$ or in terms 
of the discrete time step variable $s>L^2$.
 
\section{Analytic solution for the probability of return}
Since the boundary at $x=0$ is absorbing the first return time probability is given by the current 
$J(x,t)=-\dfrac{1}{2L}\,\dfrac{\partial }{\partial x}\left[x^a\,P(x,t)\right]$ evaluated at $x=1/L$ 
and we find 
 
\begin{equation}
\begin{split}
P_R(0,t)\simeq & \dfrac{\sqrt L}{2}\,\sum\limits_{n=1}^\infty   A_{n}\,J_{\mbox{\scriptsize 1/(2-{\it a})}} 
\Bigg(j_{\mbox{\scriptsize 1/(2-{\it a}) -1    },n}
\,{\dfrac{1}{L}}^{1-a/2}\Bigg)\\&
\times\exp\Bigg(-\dfrac{{(2-a)}^2\,j_{\mbox{\scriptsize 1/(2-{\it a}) -1    },n}^2}{8L}\,\,t\Bigg) .
\label{ReturnAnalytic0}
\end{split}
\end{equation}

\begin{figure}[ht]
 \includegraphics[scale=0.4]{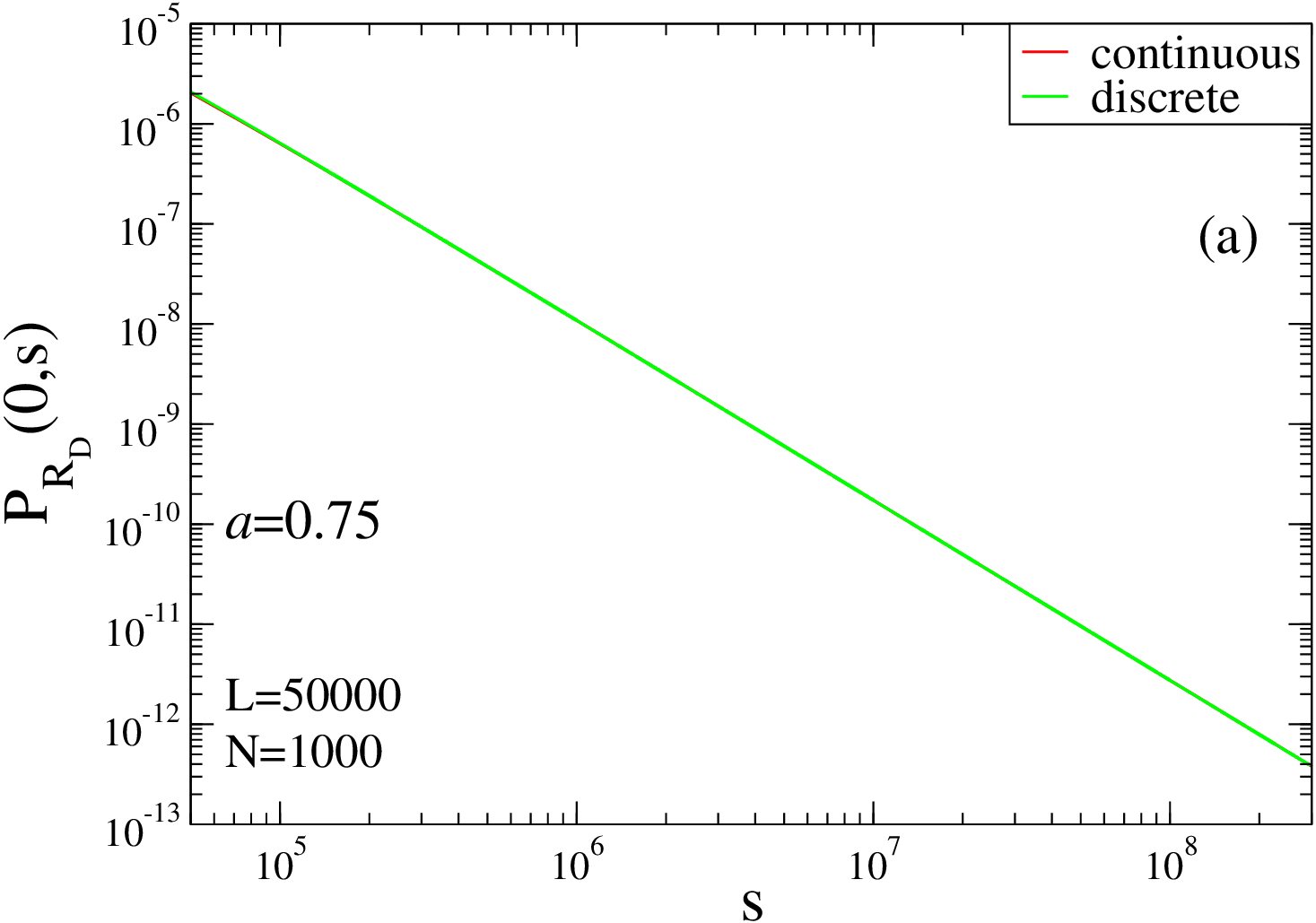}
 \includegraphics[scale=0.4]{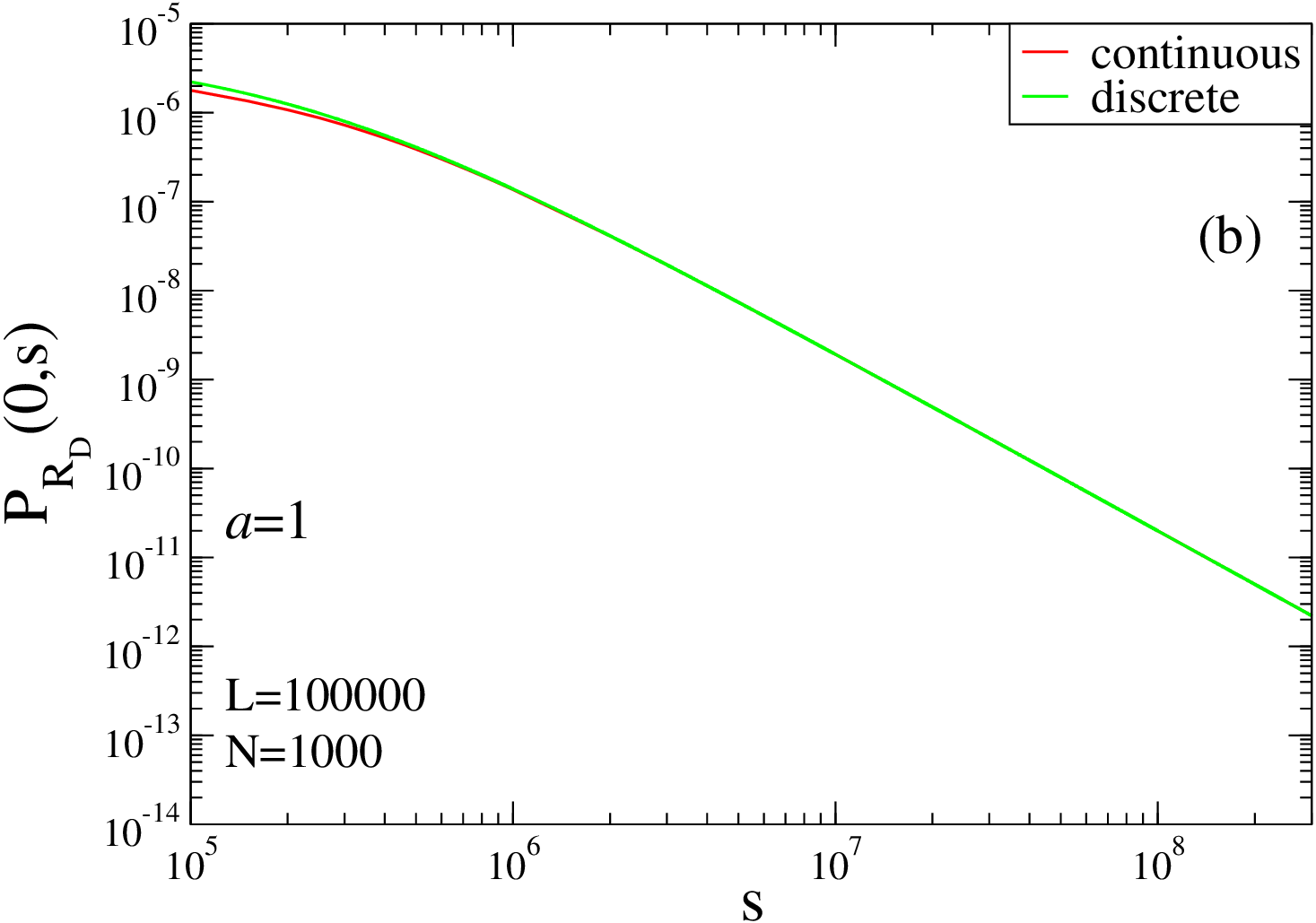}
 \includegraphics[scale=0.4]{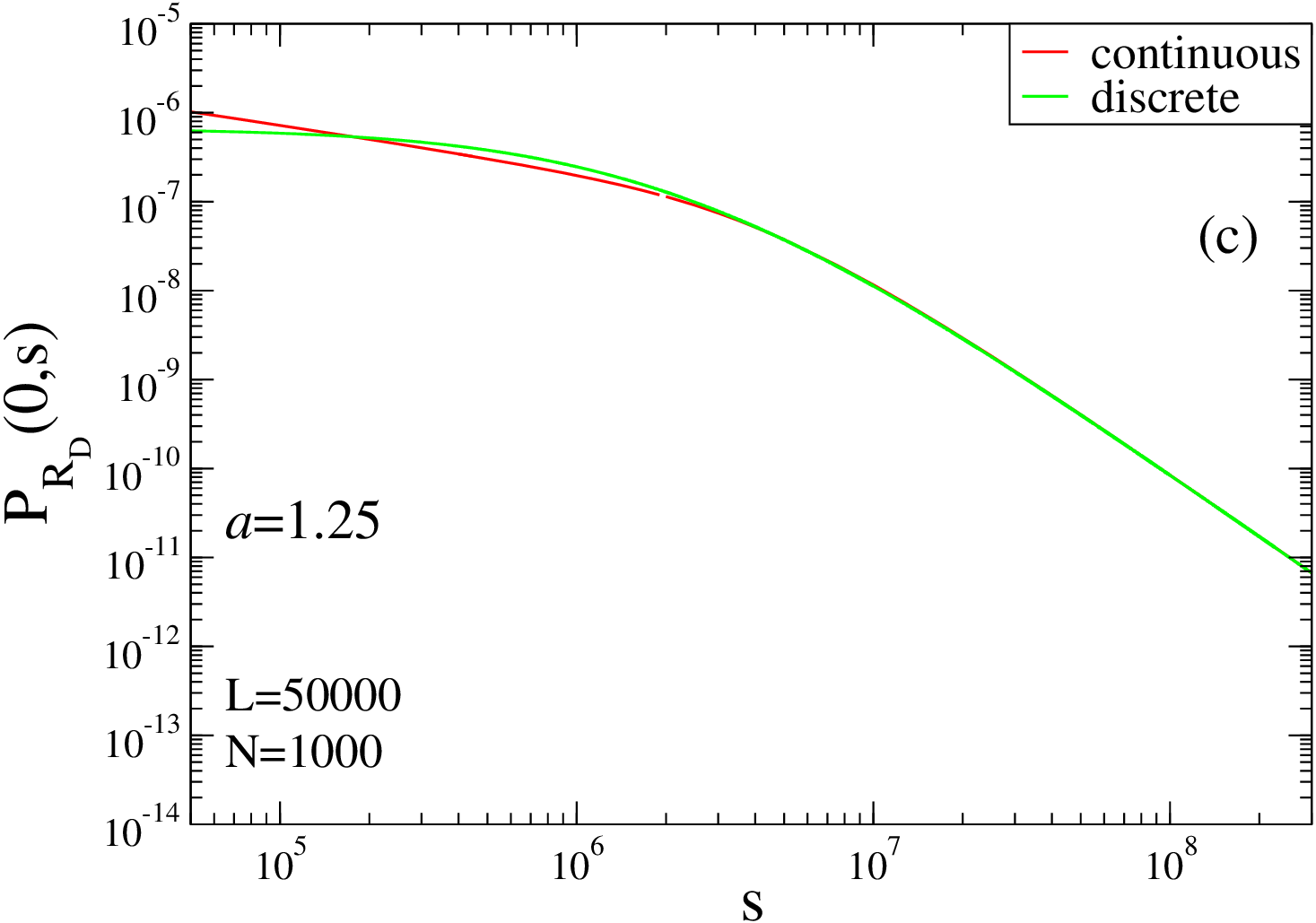}
\caption{The comparison between the discrete Master Equation (Eq.~(\ref{EqPD})) and the continuous analytic 
solution (Eq.~(\ref{ReturnAnalytic0})) for the first retrun time (as function of discrete time)for (a) $a=0.75$, (b) $a=1$ and (c) $a=1.25$.}
\label{fig1}
\end{figure}

Fig.~\ref{fig1} show a comparison between the exact numerical solution to the Master Equation 
in Eq.~(\ref{EqPD}) and the analytic solution in Eq.~(\ref{ReturnAnalytic0}). 
The number of terms is fixed to $N=1000$ which ensures that the plotted analytic result is correct 
within an accuracy smaller than the width of the plotted graph.

Unfortunately we have not been able to reduce the series in \ref{ReturnAnalytic0} to a simple 
compact expression. In particular we were unable to relate analytically $P_R(0,t)$ to the 
$q$-exponential forms given in Eq.~(\ref{qexp}). Since the Bessel functions in the series for 
$P_R(0,t)$ are smooth functions of the index $1/(2-a)$ and $1/(2-a)-1$ one will not expect 
that the description in terms of $q$-exponentials would be more appropriate for certain values 
of the exponent $a$ in Eq. (\ref{process}) than for others. However, as we'll see in the next 
section high precision numerical analysis suggests otherwise.

\section{Numerical analysis using $q$-exponentials}

Next we solve  by numerical iteration the discrete Master equation Eq. (\ref{EqPD}) to obtain 
numerically exact results for the return time distribution in Eq. (\ref{EqR}). These results 
are exact within the numerical precision of the computer.

\begin{figure}[ht]
 \includegraphics[scale=0.4]{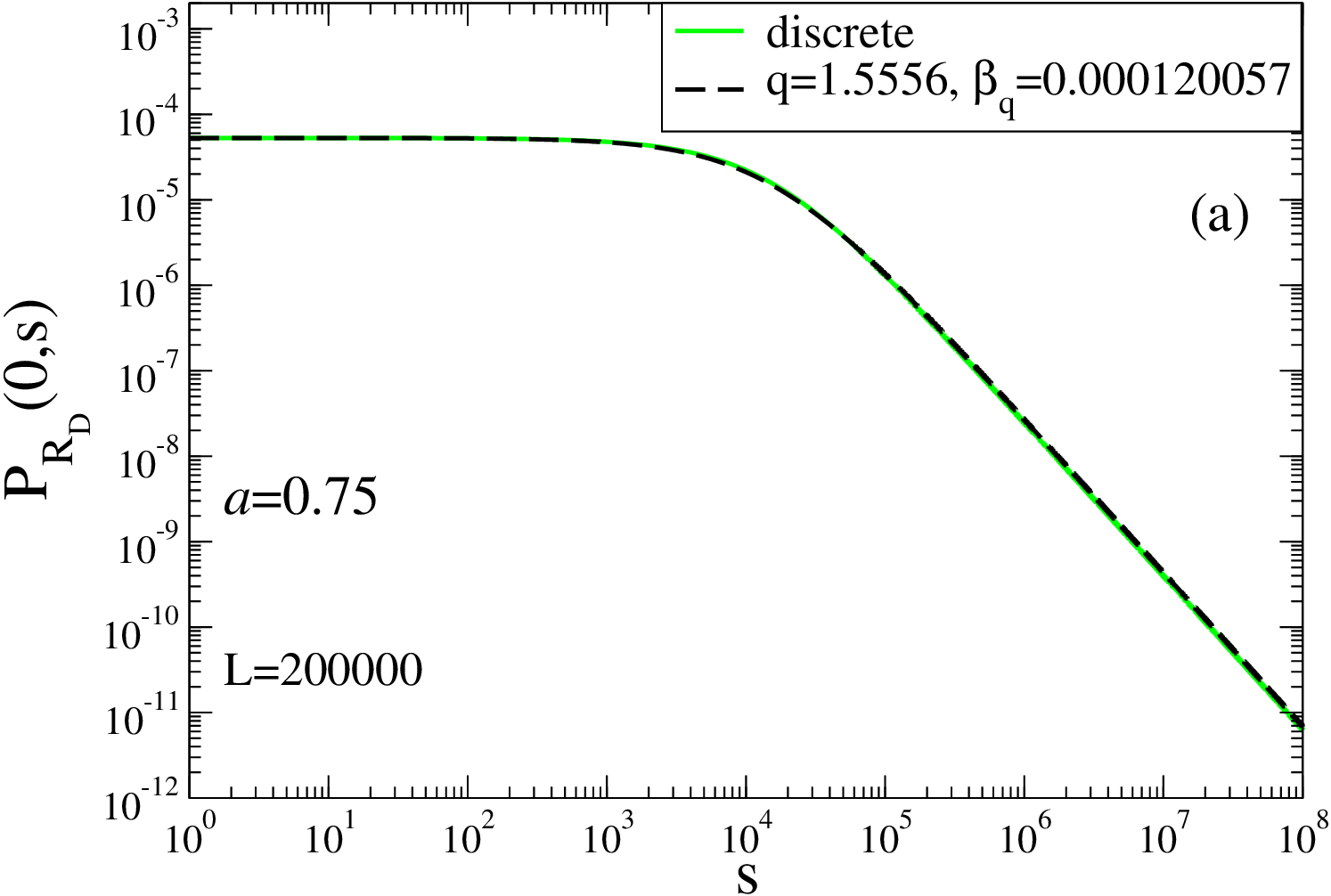}
 \includegraphics[scale=0.4]{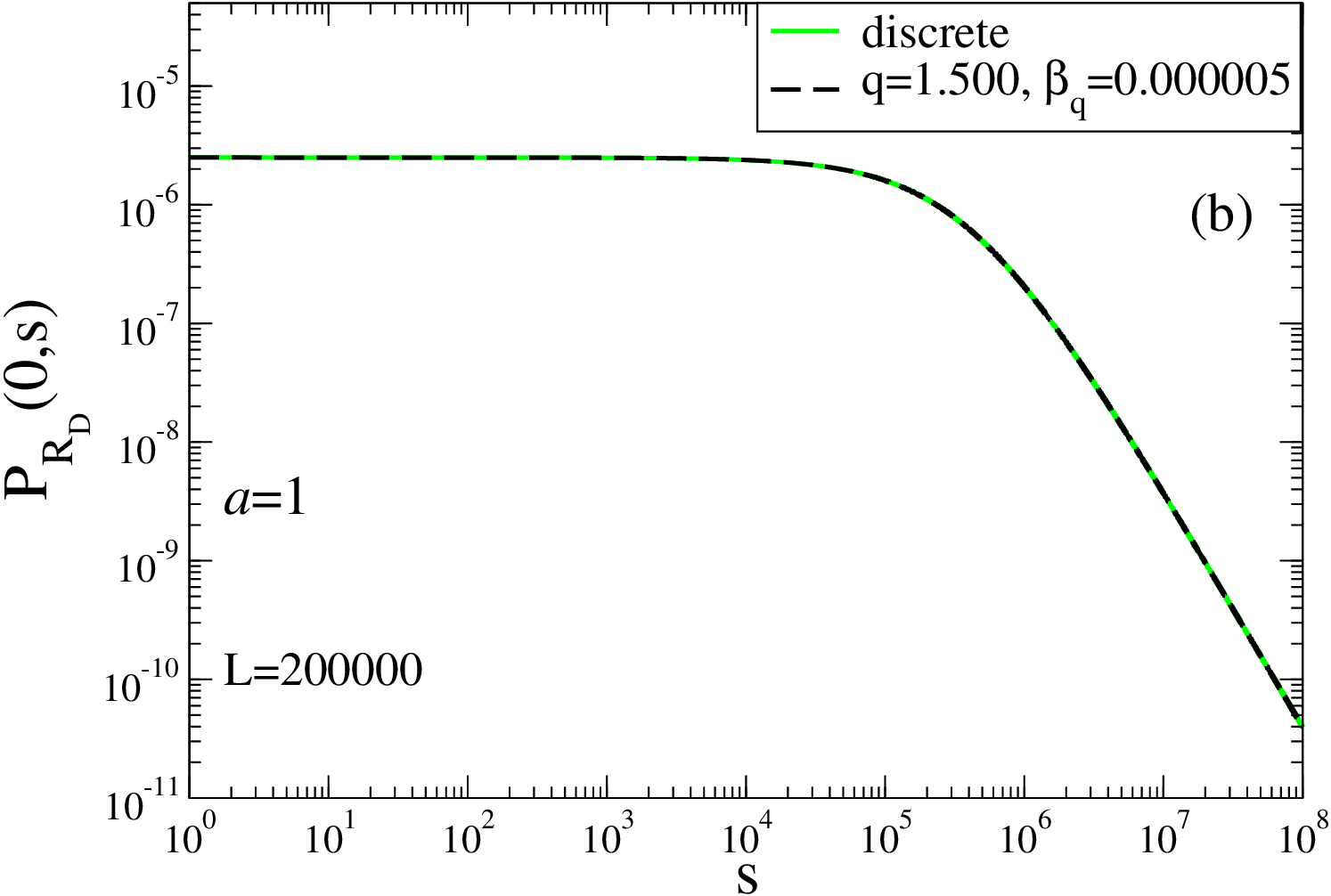}
 \includegraphics[scale=0.4]{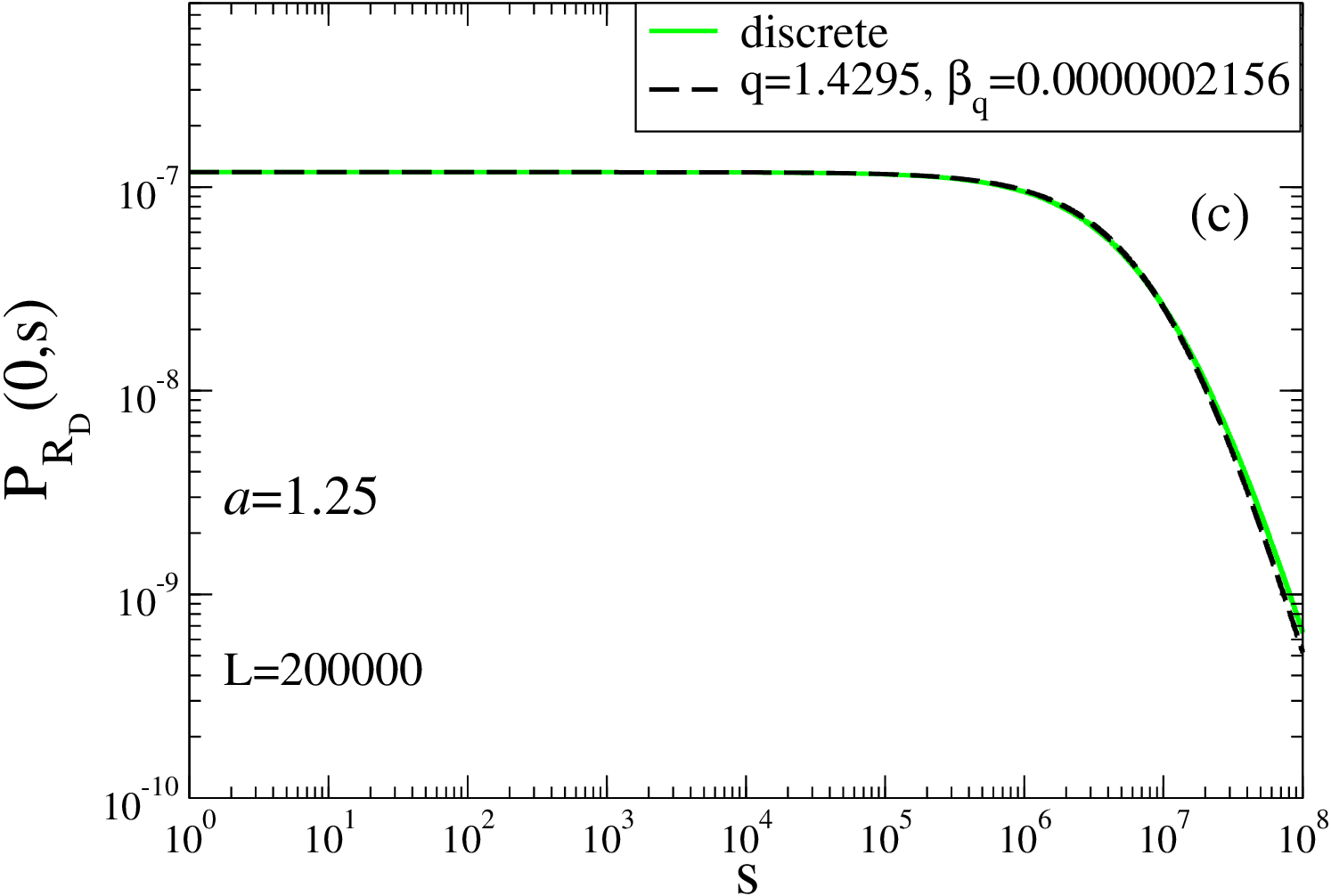}
\caption{The comparison between the discrete Master Equation (Eq.~(\ref{EqPD})) and $q$-exponential functions for the normalized first return distributions (as function of discrete time) for the cases (a) $a=0.75$, (b) $a=1$ and (c) $a=1.25$. 
  }
\label{fig3}
\end{figure}

Using the definition of $q$-exponentials in Eq.~(\ref{qexp}), one can easily write the normalized first return distributions as 

\begin{equation}
P_{R_D}\left(0,s\right) = P_{R_D}\left(0,0\right) \exp_q\left(-\beta_q s \right)_+  
\label{qexpnorm}
\end{equation}
where $\beta_q>0$ is the only free parameter to be fitted since the value of $q$ can be obtained from the asymptotic behaviour of the slope, which must be equal to $1/(q-1)$. The comparison between the discrete analytic results (Eq.~(\ref{EqPD})) and $q$-exponential (Eq.~(\ref{qexpnorm})) is given in Fig.~\ref{fig3}.  

Because the numerical study is limited to finite time the normalization will not be exact, since we lack some contribution to the return distribution from the tail of very large times. We therefore check (see Table 1) the extend to which the normalization is fulfilled for the return times we are able to handle. 

\begin{table}
\centering
{
\begin{tabular}{|c|c|}
\hline 
\hline
$L$  & normalization \\ 
\hline 
\hline 
20000& 0.9899 \\ 
\hline 
50000& 0.9899 \\ 
\hline 
100000& 0.9899 \\ 
\hline 
200000& 0.9900 \\ 
\hline 
\end{tabular}}
\caption{For various values of $L$ with $a=1$, the normalizations attained from simulations are given.}
\end{table}


A very good agreement can easily be seen from the figure, but at this point we make one more step to quantify this agreement better. To achieve this, we define a quantity $\Delta$, which is the area between the curve of the exact discrete result and related $q$-exponential function. We know that for finite $L$ the return time distribution cuts off exponential for (discrete) time larger than $L^2$. This means that the $q$-exponential form will only be valid for all times in the limit $L\rightarrow\infty$. If the distribution, in this limit, is {\it exactly} a $q$-exponential, then $\Delta$ must approach  zero as the system size tends to infinity.

\begin{figure}[ht]
 \includegraphics[scale=0.55]{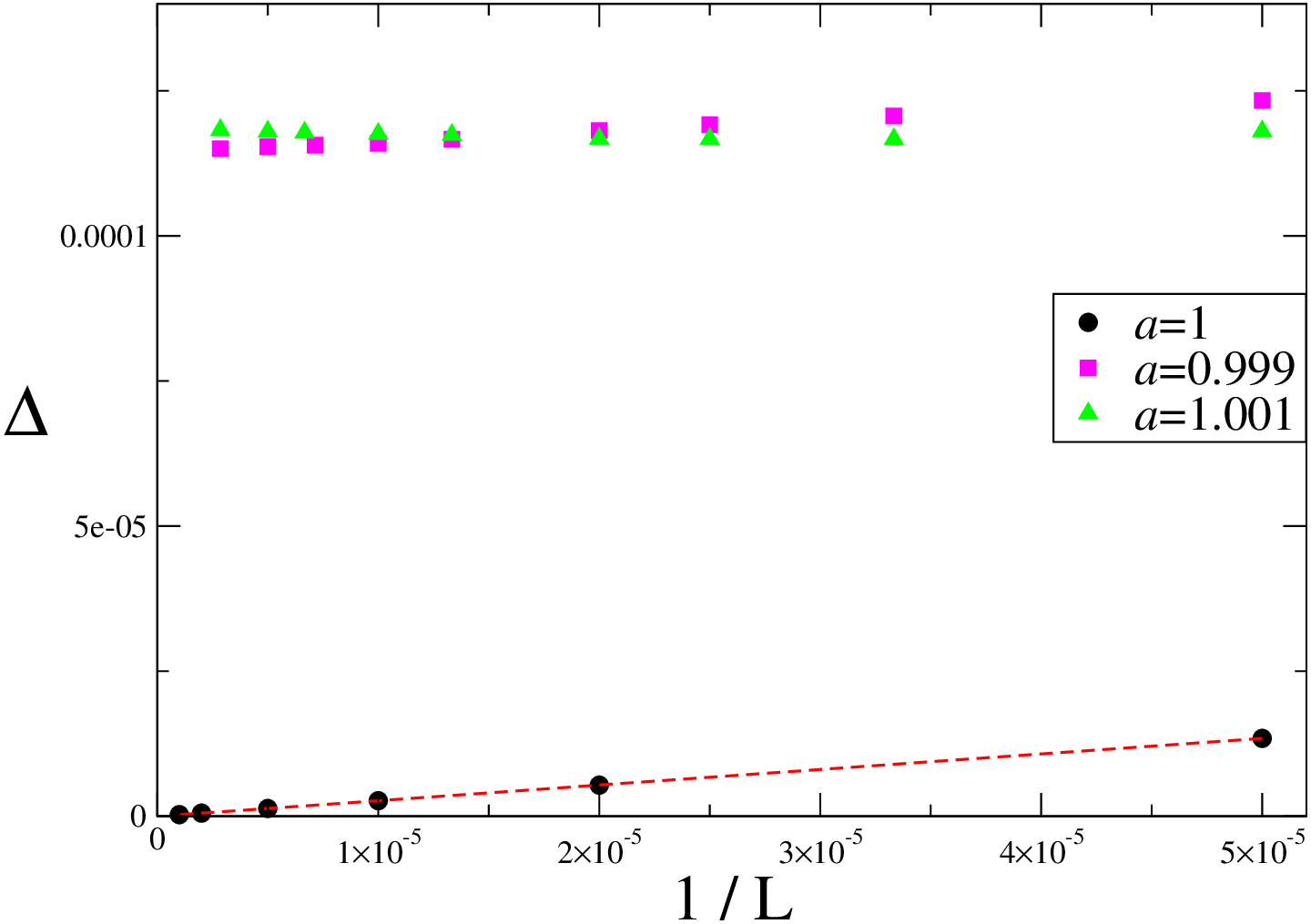}
\caption{The quantity $\Delta$ as a function of the system size.}
\label{fig4}
\end{figure}

We plot in Fig.~\ref{fig4} this quantity as a function of system size for $a=1$ case and also for two very tiny departures from this case. We note that for $a=1$ the limit $L\rightarrow\infty$ is indeed consistent with a $q$-exponential. For $a\neq 1$ by just very small amount   we obtain $\lim_{L\rightarrow\infty} \Delta>0$ and therefore strictly speaking the return distribution is not equal to a $q$-exponential for $a\neq1$. 

\section{Discussion and Conclusion}
We have analytically and numerically analysed a simple random walk with a position dependent 
transition rate with the aim to better understand when and why $q$-functional forms describe 
the statistical properties of non-conventional systems. We focus on the first return time distribution.

The analytic solution suggests that the return distributions are smoothly dependent on the $a$ parameter, 
which determines  the transition probabilities of the RRW model. But despite of being unable analytically 
to relate the return time distribution to $q$-exponentials, we found numerically that, in the limit of 
large systems sizes, the $q$-exponential fits remarkably well the numerical exact result. And indeed for 
a special value $a=1$ the numerical analysis demonstrate that the return distribution becomes equal to 
the $q$-exponential with $q=3/2$ when the system size is taken to infinity.

The question then is what is so special about $a=1$ that for this value   of $a$ the $q$-exponential is 
{\em equal} to the considered distribution and not just a very good fit?

One can identify a few ways in which $a=1$ is special. The mean of the return time can be calculated 
directly (see e.g.\cite{Pavliotis2014} and \cite{Redner2001}) without the need of knowing 
the entire return time distribution. The average of the number of discrete time 
steps for the first return from position $n=1$ to $n=0$, which we denote by $\langle s_R\rangle$,  is to leading order in the system size $L$ given by

\[	
	\langle s_R\rangle =\left\{\begin{array}{ll}
	                    L/(1-a) & \mbox{for $0<a<1$} \\
	                    L\ln L  & \mbox{for $a=1$} \\
	                    L^a/(a-2) & \mbox{for $1<a<2$.} 
                        \end{array}
	                    \right.
\]
This indicates one way in which $a=1$ is marginal.  

The $q$-exponential found for $a=1$ has index $q=3/2$ which corresponds to a functional form 
$f(x) = a/(1+x)^2$. We want to mention that this form is also the functional dependence of the 
distribution of extinction times for a critical birth-death process in which the rate of death is 
equal to the rate of birth \cite{Pruessner_private}. This result is likely to be related to the RRW 
studied here when $a=1$, since the birth-death process is related to a critical branching process, 
which on the other hand can be related to a random walk with a transition probability proportional 
to the position \cite{Pruessner2012}.

Given these remarks it is difficult to decide whether the success of the fit to $q$-exponentials is 
caused by some fundamental underlying principle or is accidental in origin.

\section*{Acknowlegment}
HJJ is greatful for very helpful and insightful discussions with Gunnar Pruessner and Grigoris A Pavliotis. 
This work has been supported by TUBITAK (Turkish Agency) under the Research Project number 112T083. 
U.T. is a member of the Science Academy, Istanbul, Turkey.


\end{document}